\documentclass[twocolumn,prl,epsfig,aps]{revtex4-1}

\usepackage{graphicx,amsmath}

\newcommand{\beq}{\begin{eqnarray}}
\newcommand{\eeq}{\end{eqnarray}}

\begin{document}
\setcounter{page}{1}

\title{Superfluid and supersolid phases of $^{\bf 4}$He on the second layer of graphite 
}
\author{M.C. Gordillo}
\affiliation{Departamento de Sistemas F\'{\i}sicos, Qu\'{\i}micos y Naturales,
Universidad Pablo de Olavide. E-41013 Seville, Spain}
\author{J. Boronat}
\affiliation{Departament de F\'{\i}sica, 
Universitat Polit\`ecnica de Catalunya, 
Campus Nord B4-B5, E-08034 Barcelona, Spain}

\begin{abstract} 
We revisited the phase diagram of 
the second layer of $^4$He on top of graphite using quantum Monte Carlo 
methods. Our aim was to explore the existence of
the novel phases suggested recently in experimental works, and determine 
their properties and stability limits. We found evidence of a superfluid 
quantum phase with hexatic correlations, induced by the corrugation of the 
first Helium layer, and a quasi-two-dimensional supersolid corresponding to a 
7/12 registered phase. The 4/7 commensurate solid was found to be unstable, 
while the triangular incommensurate crystals, stable at large densities, 
were normal.        
\end{abstract}


\maketitle

The light mass of Helium atoms and the strong Carbon-Helium interaction make 
$^4$He adsorbed on graphite the most paradigmatic example of a two-dimensional 
(2D) quantum system. 
Its phase diagram was extensively studied in the 90’s, 
using a variety of experimental techniques 
(see, for instance, Ref. \onlinecite{cole}). The consensus so far is  
that, at very low temperature, $^4$He in direct contact with the graphite 
surface is a $\sqrt3 \times \sqrt3$ registered solid 
that undergoes a first-order phase transition to a incommensurate 
triangular 2D crystal upon increasing the Helium density. This was also 
confirmed by first-principles theoretical descriptions of the 
system~\cite{corboz,yo}. Quantum Monte Carlo simulations in the limit of 
zero temperature found other proposed commensurate phases to be 
unstable~\cite{yo}.  

By increasing the Helium coverage, the system undergoes first-order layering 
transitions, a feature that was clearly observed recently on a single 
carbon nanotube~\cite{adrian}. 
In graphite, there seemed to be a consensus about the second 
$^4$He layer, stable in the coverage range $\sim$ 0.114-0.200 \AA$^{-2}$. Those are total densities, including
helium atoms per surface unit both in the first and second layers. 
Heat capacity~\cite{greywall1,greywall} 
and torsional oscillator~\cite{B7,B8} experiments indicated that, after a 
promotion from the first to the second layer 
a quasi-two-dimensional liquid was formed. 
Increasing the coverage, the liquid changes into a commensurate 
phase (with respect to the first adsorbed Helium layer), and then to an 
incommensurate one before promotion 
to a third layer~\cite{greywall1,greywall}. 

Remarkably, two recent experimental works have reopened the doubts about the phase diagram of the second layer of $^4$He adsorbed on graphite.  
First, the calorimetric data of Ref. \onlinecite{B12} suggests the existence of a liquid above 0.175 \AA$^{-2}$, followed upon an increasing of the 
helium coverage, of a stable phase in the 0.196-0.203 \AA$^{-2}$ range. That phase could be either a commensurate solid {\em or} 
a quantum hexatic phase. 
On the other hand, the torsional oscillator data of 
Ref. \onlinecite{B13} indicate a normal 2D liquid between $\sim$0.1657 \AA$^{-2}$ and 
0.1711 \AA$^{-2}$ (Ref. \onlinecite{B13}, supplementary information), followed by 
an arrangement showing a superfluid response from 0.1711 to 0.1996 
\AA$^{-2}$, with a maximum at 0.1809 \AA$^{-2}$.
Since, according to previous DMC calculations \cite{B10}, those densities are above the stability limits of a liquid phase, that response would correspond 
to a quasi-two-dimensional supersolid. This would 
be the first indication of a stable supersolid phase in $^4$He, after 
discarding that possibility in bulk~\cite{chan}.

In this Letter, we revisit this problem from a theoretical microscopic point of 
view. Our aim is to clarify the nature of the stable phases of the 
second layer of $^4$He on graphite, in the limit of zero temperature. Our 
results show hexatic order~\cite{hexatic,hexatic2,hexatic3} before 
crystallization into one of the possible registered phases (7/12). In both 
cases, our measure of the superfluid fraction gives a finite value, larger for 
the hexatic but still very significant for the registered solid. Therefore, on this 
layer we found two long pursued phases: a superhexatic~\cite{superhexatic} 
and a quasi-two dimensional (registered) supersolid.

Our zero-temperature first-principles study relies on the diffusion Monte Carlo 
(DMC) method, used extensively in the past to analyze $^4$He phases 
in different geometries~\cite{borobook}. The high numerical accuracy of DMC in 
the estimation of the energy is crucial to disentangle the stability of 
different possible phases. This is specially relevant in the study of  
registered phases since the energy differences between different 
commensurate solids is very tiny.
In essence, the DMC method allows us to solve           
the many-body imaginary-time Schr\"odinger equation corresponding to the 
Hamiltonian describing the system~\cite{boro94}. In the present case, 
\begin{equation} \label{hamiltonian}
H = \sum_{i=1}^N  \left[ -\frac{\hbar^2}{2m} \nabla_i^2 + 
V_{{\rm ext}}(x_i,y_i,z_i) \right] + \sum_{i<j}^N V_{\rm{He-He}} 
(r_{ij}) \ ,
\end{equation}
where $x_i$, $y_i$, and $z_i$ are the coordinates of the $N$ Helium atoms, 
including first and second layers, and $m$ is the $^4$He mass.
Graphite was modeled by a set of eight graphene layers separated $3.35$ \AA$ $ 
in the $z$ direction and stacked in the A-B-A-B way typical of this compound, as in Ref. \onlinecite{yo}. 
We stopped at eight graphene sheets since to include a ninth one changes the energy per particle less than 
the typical error bars for that magnitude (around 0.1 K) \cite{yo}.
$V_{{\rm ext}}(x_i,y_i,z_i)$ sums, for each $^4$He atom $i$, all the  C-He pair 
interactions  calculated using the accurate Carlos 
and Cole anisotropic potential~\cite{carlosandcole}. The He-He 
interaction $V_{{\rm He-He}} (r_{ij})$ is modeled using a standard
Aziz potential~\cite{aziz}. The graphite substrate
was considered to be rigid, but the helium atoms in both the first and second layers were allowed to move from their crystallographic positions.

In order to reduce the variance and to fix the phase 
under study, DMC incorporates importance sampling by using a guiding wave 
function. This wave function is designed as a first, simple approximation 
to the many-body system and is variationally optimized. In 
our case, we used
\begin{eqnarray}
\Phi({\bf r}_1,\ldots,{\bf r}_N) & = & \Phi_J({\bf r}_1,\ldots,{\bf r}_N)
\Phi_1({\bf r}_1,\ldots,{\bf r}_{N_1})  \nonumber \\
& & \times \ \Phi_2({\bf r}_{N_1+1},\ldots,{\bf r}_N) \ ,
\label{phitot}
\end{eqnarray}
with 
\begin{equation}
\Phi_J({\bf r}_1,\ldots,{\bf r}_N) = \prod_{i<j}^{N} \exp \left[-\frac{1}{2} 
\left(\frac{b}{r_{ij}} \right)^5 \right] \ 
\label{sverlet}
\end{equation}
a Bijl-Jastrow wave function built as a product of  McMillan pair correlation 
factors with a variational parameter $b$, whose value was taken from 
the literature~\cite{boro94,yo}. The one-body terms of the $N_1$ atoms in the 
first layer (\ref{phitot}) are given by 
\begin{eqnarray}
\lefteqn{\Phi_1({\bf r}_1,\ldots,{\bf r}_{N_1})  =  \prod_{i=1}^{N_1}
\Psi_1(z_i)} \nonumber \\ 
& & \times \prod_{i,I=1}^{N_1} \exp \{-a_1 [(x_i-x_I^{(1)})^2 +
(y_i-y_I^{(1)})^2] \} \ . 
\label{first} 
\end{eqnarray} 
Following Ref.~\onlinecite{yo}, $\Psi_1(z_i)$ is the solution to the one-body 
Schr\"odinger equation that describes a single Helium atom in a z-averaged 
helium-graphite potential, $a_1$
being a variational parameter different for each $^4$He first layer density. The coordinate set $(x_I^{(1)},y_I^{(1)})$ corresponds to the 
$N_1$ crystallographic sites of that first layer. On the other hand, the second
layer was described by a symmetric Nosanow function to allow for 
possible exchanges in the crystal~\cite{yo3},
\begin{eqnarray}
\lefteqn{\Phi_2({\bf r}_{N+1},\ldots,{\bf r}_{N})  =  \prod_{i=1}^{N_1}
\Psi_2(z_i)} \nonumber \\
& & \times \prod_{i,I=1}^{N_2} \left[ \sum_{i=N_1+1}^{N} \exp \{-a_2 [(x_i-x_I^{(2)})^2 +
(y_i-y_I^{(2)})^2] \} \ \right],
\label{second} 
\end{eqnarray}
Here, $\Psi_2(z_i)$ is a Gaussian of the type $\exp [-c_2 (z_i-z_m)^2]$, with 
both $c_2$ and $z_m$ variationally optimized parameters. $N_2$ stands both
for the number of Helium atoms on the second layer and for the number of 
lattice points of the solids. Therefore, no vacancies were 
considered in any solid. $a_2$ was chosen to minimize the 
total energy of the system. Notice that Eq. (\ref{second}) can describe a 
translationally invariant Helium second layer by fixing $a_2$ to zero.  
In most cases, $N_1$ was fixed to 224 atoms, distributed in a simulation box 
including $14\times 8$  triangular unit cells, while $N_2$ was fixed to 
produce the desired density. This meant simulation cells including up to 356 
$^4$He atoms. 

\begin{figure}
\begin{center}
\includegraphics[width=0.8\linewidth]{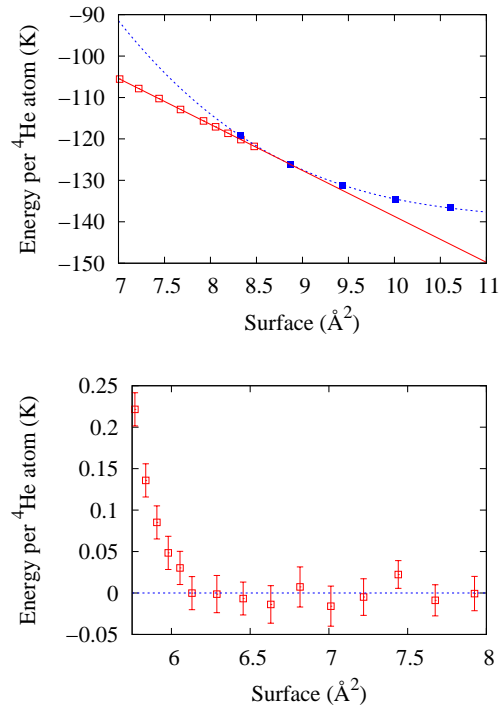} 
\caption{
{\em Top panel:} Full squares: Energy per $^4$He atom in a single layer of a 
triangular solid as a function of surface per helium atom. 
Open squares: Same for an arrangement comprising two helium layers, the density of the first one being 0.115 \AA$^{-2}$. A least-squares fitting polynomial  
to the first layer data and the Maxwell construction line between the two arrangements are also shown.                
{\em Bottom panel:} Data for the energy of the second layer system after 
subtracting the values given by the Maxwell construction line.   
}
\label{fig1}
\end{center}
\end{figure}

\begin{figure}
\begin{center}
\includegraphics[width=0.8\linewidth]{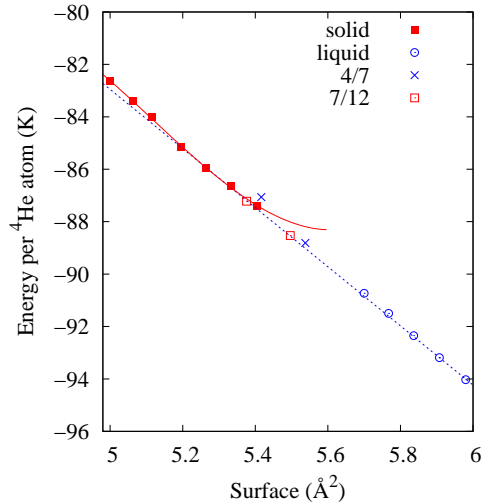} 
\caption{ 
Same than in the previous figure for the two layer system at higher densities. 
Data for different phases are displayed.   
}
\label{fig2}
\end{center}
\end{figure}

Calculating the energy per particle as a function of density, one can establish 
the stability range of the different phases. The DMC energies are 
shown in Figs. \ref{fig1} and \ref{fig2} for different values 
of the surface area (the inverse of the surface density). 
The first issue to be addressed is the determination the  first-layer solid 
density which produces the lowest total energy for a 
two-layer arrangement. To this end, we followed a procedure used previously for 
graphene~\cite{B10}.
As in that work, the 
optimum density of the first layer turned out to be 0.115 \AA$^{-2}$. A 
standard Maxwell 
construction between a system with a single Helium monolayer and a 
double-layered one with that solid density 
(Fig. \ref{fig1}, top panel) and a translational invariant second layer, gives 
us  a promotion density of 0.113 $\pm$ 0.002 \AA$^{-2}$ (corresponding to
8.86 $\pm$ 0.15 \AA$^2$). This is similar to the experimental value 0.114 
\AA$^{-2}$ reported in Ref. \onlinecite{B13} and somewhat smaller than the 
other experimental measure, 0.118 \AA$^{-2}$, that of Ref. \onlinecite{B12}. 
To calculate the lowest density limit for the bilayer arrangement, we 
subtracted the energy values for the Maxwell construction line
from the direct simulation results. The results obtained are shown in the 
bottom panel of Fig. \ref{fig1}. As one can see, this limiting value is 
6.13 $\pm$ 0.15 \AA$^2$, which corresponds to a density $\sigma = 0.163 \pm 
0.002$ \AA$^{-2}$.

A similar procedure allows us to establish the stability limits for 
larger values of the total $^4$He density. The results are depicted in 
Fig.~\ref{fig2}. There, we can see the energy per 
atom for different phases. For total density values smaller than 0.185 
\AA$^{-2}$, the underlying Helium density was, as before, 0.115 \AA$^{-2}$, 
while for that value up, the energy 
was lowered by slightly compressing the first layer to a density 
 0.1175 \AA$^{-2}$. In Fig.~\ref{fig2}, the open circles correspond to 
the same translationally invariant double layer arrangement already discussed 
(Fig.~\ref{fig1}). 
The full squares correspond to a bilayer comprising two 
incommensurate triangular solid layers. 
The 4/7 and 7/12 commensurate phases with the 
triangular lattice of the first layer         
were also considered by an appropriate choice of the set of crystallographic 
positions defining them. The fact that, in the figure, there are two points for 
each of those arrangements is due to the fact that 
we considered the two possibilities for the first-layer densities discussed 
above. In any case, it is clear from Fig. \ref{fig2}, that the 4/7 arrangement 
is unstable. On the other hand, the 7/12 phase is right
on top of the Maxwell construction line between the translationally-invariant 
phase and the double triangular solid one. This means we would have a first 
order phase transition between a translationally-invariant 
phase of density $\sigma = 0.170 \pm 0.002$ \AA$^{-2}$, and a 7/12 
registered solid of 0.182 \AA$^{-2}$. The first layer of this arrangement would 
be then compressed up to  $\sigma = 0.186$ \AA$^{-2}$, that is again 
in equilibrium with a triangular solid of density 0.188 \AA$^{-2}$. This is 
similar to what happens in a $^3$He double layer system on graphite~\cite{yo5}, 
where both the 4/7 and 7/12 phases are stable.

\begin{figure}
\begin{center}
\includegraphics[width=0.8\linewidth]{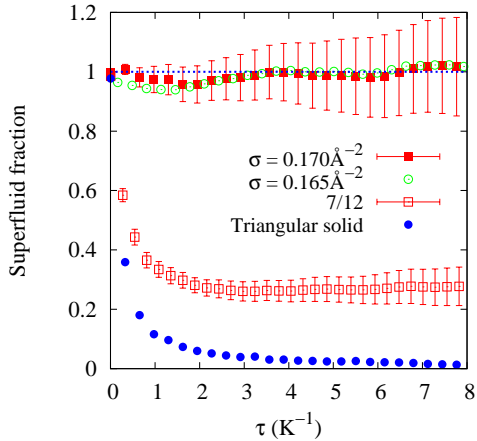} 
\caption{Superfluid 
density for different second layer arrangements. Full squares, a 
translationally-invariant phase with $\sigma=0.170$ \AA$^{-2}$. Open circles, 
same but for $\sigma= 0.165$ \AA$^{-2}$. Open squares, results for a 7/12 phase 
with $\sigma = 0.186$ \AA$^{-2}$. Full circles, same for a triangular solid 
with $\rho =0.196$ \AA$^{-2}$. The dotted line corresponds to a perfect 
superfluid with no normal component.        
}
\label{fig3}
\end{center}
\end{figure}

To  better characterize the different stable phases, we need to go further than 
to establish their stability limits. As indicated above, one of the issues 
raised in Ref. \onlinecite{B13} was the possible
existence of a quasi-two-dimensional supersolid and of a normal (non 
superfluid) liquid phase at smaller densities. In order to study those 
claims, we estimated the superfluid fraction $\sigma_s/\sigma$ on the second 
layer of the arrangements found to be stable. 
This is done by using the usual 
winding-number estimator in the limit of zero temperature~\cite{yo3,gubernatis},
\begin{equation}
\frac{\sigma_s}{\sigma}= \lim_{\tau \rightarrow \infty} \alpha \left( 
\frac{D_s(\tau)}{\tau} \right) \ ,
\end{equation}
with $\tau$ the imaginary time in the Monte Carlo 
simulation. Here, $\alpha = N_2/(4 D_0)$, $D_0 = \hbar^2/(2m)$, and 
$D_s(\tau) = \langle [{\bf R}_{CM}(\tau)-{\bf R}_{CM}(0)]^2 \rangle$. ${\bf 
R}_{CM}$ is the 
position of the center of mass of the $N_2$ $^4$He atoms on the second layer, 
taking into account only their $x$ and $y$ coordinates (where periodic 
boundary conditions apply). The results obtained for different cases are shown 
in Fig.~\ref{fig3}.    
The error bars in that figure correspond to one standard deviation computed 
from several independent Monte Carlo runs, and when not displayed, are similar 
to those shown.

In Fig.~\ref{fig3}, we see that the translationally 
invariant phase akin to a liquid found at low densities is a perfect superfluid, 
since  
the values for the superfluid estimator are on top to the line corresponding to 
$\sigma_s/\sigma=1$. On the other hand, the superfluid estimator for a 
triangular solid of $\sigma=0.196$ \AA$^{-2}$ is zero 
within our numerical resolution . This situation is common to all other 
triangular lattice solids, irrespective of their total density, 
and makes them normal solids.       
However, the superfluid fraction corresponding to a 7/12 structure with
$\sigma=0.186$ \AA$^{-2}$ (open squares) has an intermediate value between zero 
and one, 
corresponding to a superfluid fraction of $0.3 \pm 0.1$. This fraction is the 
same as the one for a $\sigma=0.182$ \AA$^{-2}$ solid, whose data are not shown 
for simplicity.  
This means there is a supersolid stable phase in the density range 
0.182-0.186\AA$^{-2}$. Our results also support the existence of superfluidity 
in the range between 0.170 and 0.182 \AA$^{-2}$,
in which there is a mixture of a full superfluid liquid-like phase in 
coexistence with the 7/12 registered solid with the lowest density. The same can 
be said of the 0.186-0.188\AA$^{-2}$ interval, where the coexistence is with a 
normal triangular 2D-solid.

\begin{figure}
\begin{center}
\includegraphics[width=0.8\linewidth]{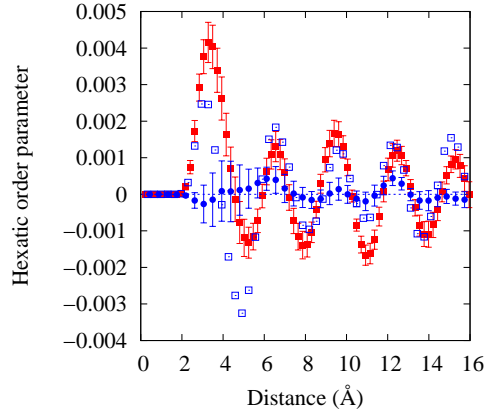} 
\caption{  
Hexatic order parameter for translationally invariant phases of $\sigma= 
0.170$ \AA$^{-2}$ (full squares) and $\sigma= 0.163$ \AA$^{-2}$ (open squares). 
Full  circles correspond to a phase with $\sigma = 0.165$ \AA$^{-2}$, but 
in which the effect of the first solid layer have been smoothed out.   
When not displayed, the error bars are similar to the ones shown. 
} 
\label{fig4}
\end{center}
\end{figure}

Previous studies on similar systems have always assumed the translationally-invariant phase to be an homogeneous liquid. In this work, we have checked that assumption by considering the 
possibility of the second layer being some kind of hexatic phase, 
as suggested in Ref. \onlinecite{B12}. To study if this was so, and following 
Ref. \onlinecite{apaja}, we write the pair distribution function $g(r)$ as
\begin{equation}
g(r) = \sum_{n=0}^{\infty} g_n(r) \cos( n \theta)
\end{equation}     
with $n$ even and $\theta = {\bf r} \cdot {\bf e}_0$. Here, ${\bf e}_0$ is a 
unit vector in a reference direction. The hexatic order in a non-solid phase is 
associated to a periodic oscillation in the $g_6(r)$ component with algebraic 
decay at large distances~\cite{hexatic,hexatic2,hexatic3,apaja}. 
This hexatic order parameter is shown in Fig.~\ref{fig4} for two translational 
invariant structures within 
the stability range of that phase. It is important to stress that the guiding 
wave function  used in DMC to describe those arrangements, a product of Eqs. 
(\ref{sverlet}), (\ref{first}) and (\ref{second}), 
this last one with $a_2 = 0$, does not include any explicit hexatic correlation. 
What we see is a regular pattern of maxima and minima        
extending to long distances with a slow decay that we cannot determine 
completely due to the finite size of our simulation box. This ordering is 
similar to the one found to be metastable in  strictly 2D $^4$He at larger 
densities ($\sigma > 0.060$  \AA$^{-2}$)~\cite{apaja} that the second-layer 
ones for the systems
under consideration (0.047 and 0.055 \AA$^{-2}$). This fact
suggests it to 
be  a consequence of the corrugation of the
first layer solid substrate. This underlying structure supports a series of 
potential minima between every three Helium atoms with the right symmetry to 
produce the observed order. To check if this is so, 
we calculated the same parameter for a structure in which the potential created by the second layer had been averaged over to make those potential minima disappear. The result is that the 
set of maxima and minima are not longer present, indicating that the hexatic correlations are indeed corrugation-induced.

Our theoretical results are at least qualitatively compatible  with the 
available experimental data. For instance, both Refs. \onlinecite{B12} and 
\onlinecite{B13} suggest that the 
phase diagram of the second layer of $^4$He on graphite starts right after 
promotion with a gas-liquid coexistence zone, followed upon an increase of 
Helium density to a stable liquid-like
region. This would be, at least for the lowest densities ($\sigma < 0.170$ 
\AA$^{-2}$), a normal fluid that would undergo a first-order phase transition 
to a commensurate phase. This last 
phase would change to a high-density triangular solid. Ref. \onlinecite{B13} assigns a density range of 0.1711-0.1809 \AA$^{-2}$ to the liquid-commensurate transition, and  
of 0.1809-0.1841 \AA$^{-2}$ for the stable registered phase. Both of them are 
comparable with our suggestions: 0.170-0.182 \AA$^{-2}$ and 0.182-0.186 
\AA$^{-2}$, while the data 
of Ref. \onlinecite{B12} is shifted further up in the density scale. In that 
entire range, we see a superfluid response, first in the coexistence between the 
superhexatic~\cite{superhexatic}
phase and the 7/12 registered solid, and then in the stability range of that 
phase itself, in accordance with the experimental data of both Refs. 
\onlinecite{B8} and \onlinecite{B13}. 
On the other hand, previous DMC calculations on graphene  
found the 7/12 commensurate solid to be unstable \cite{B10}. This difference in the behavior of those close related systems
can be due to the delicate energy balance needed for that structure to be seen: the introduction of the exchanges in the description of the supersolid decreases the 
energy enough for this commensurate phase to emerge at $T=0$ K. The effect of 
the additional carbon layers might also play a role, as in the case of the first-layer $\sqrt3 \times \sqrt3$ phase, which is more stable with respect to the metastable liquid 
in graphite than in graphene \cite{yo}.  
On the other hand, the fact that we do see a superfluid response in the range 0.163-0.170 
\AA$^{-2}$ instead of the normal fluid found experimentally, can be due to the lack of connectivity of 
the real substrate \cite{B8}.  
Finally, the lack of that response beyond 0.188 \AA$^{-2}$, is in agreement with the results of Ref. \onlinecite{B8}. 

Our work is carried out strictly at zero temperature and compares very well 
with recent experiments performed in the mK regime. It would be very 
interesting to study the same system at finite temperature to determine the 
thermal stability of the superhexatic and supersolid phases, even though those phases could be
unstable at temperature values too low to be accessible 
to the path integral Monte
Carlo method~\cite{corboz}.

\acknowledgments
We acknowledge financial support from  
MINECO (Spain) Grants FIS2017-84114-C2-2-P  and FIS2017-84114-C2-1-P.
We also acknowledge the use of the C3UPO
computer facilities at the Universidad Pablo de Olavide.


\end{document}